\documentclass[twocolumn,prb,aps,showpacs,10pt]{revtex4-1}
\usepackage{amsfonts}
\usepackage{amsmath}
\usepackage{amssymb}
\usepackage{graphicx}
\usepackage{bm}
\usepackage{color}

\newcommand{\K}[6]{\bar K_{\sigma{#1}  n{#3} {\bf k}{#5}}^{\sigma{#2}  n{#4} {\bf k}{#6} }}

\newcommand*{\M}[2]{M_{n{#1}}^{n{#2}}}

\newcommand*{\Ck}[4]{\bar C_{\sigma{#1} {\mathbf k}{#3}}^{\sigma{#2} {\mathbf k}{#4}}}
\newcommand*{\C}[3]{C_{\sigma{#1} {\mathbf k}{#3}}^{\sigma{#2}  }}

\newcommand*{\dt}{\frac{\partial}{\partial t}}
\newcommand*{\dkd}{d^D{\mathbf k}}
\newcommand*{\bk}{{\mathbf k}}
\newcommand*{\bs}{{\mathbf s}}
\newcommand*{\bS}{{\mathbf S}}

\newcommand*{\bR}{{\mathbf R}}
\newcommand*{\nn}{\nonumber \\}

\begin{document}

\pacs{75.78.Jp, 75.50.Pp, , 78.47.J-, 75.30.Hx}

\title{Non-Markovian  spin transfer dynamics in magnetic semiconductors despite short memory times}

\author{C. Thurn}
\author{M. Cygorek}
\author{V. M. Axt}
\affiliation{Theoretische Physik III, Universit{\"a}t Bayreuth, 95440 Bayreuth, Germany}
\author{T. Kuhn}
\affiliation{Institut f\"ur Festk\"orpertheorie, Universit\"at M\"unster, 48149 M\"unster, Germany}

\begin{abstract}
A quantum kinetic theory of the spin transfer between carriers and Mn atoms
in a Mn doped diluted magnetic semiconductor is presented.
It turns out that the typical memory time associated with these processes
is orders of magnitude shorter than the time scale of the spin transfer.
Nevertheless, Markovian rate equations, which are obtained by neglecting the
memory, work well only for bulk systems. For quantum wells and wires
the quantum kinetic results qualitatively deviate from the Markovian limit
under certain conditions. Instead of a monotonic decay of an initially prepared 
excess electron spin, an overshoot or even coherent oscillations are found.
It is demonstrated that these features are caused by energetic redistributions of
the carriers  due to the energy-time uncertainty.
\end{abstract}
\maketitle

\section{Introduction}

Ultrafast dynamics of magnetic semiconductors are a vastly growing field. 
Most studied are manganese doped diluted III-V or II-VI semicondutors and their nanostructures. \cite{Kacman:01,Wu:10}
Driven by the vision of spintronic applications, much of the research 
is devoted towards the control of magnetic properties on ultrashort time scales.
Here, optical methods are particularly attractive. 
Experiments as well as theoretical studies
have revealed that optical manipulation of the magnetization
is possible in a coherent non-thermal regime\cite{crooker:97,wang:09,Kapetanakis:09,Qi:09,Reiter:09,Reiter:11,Reiter:12,Kirilyuk:10} 
which opens many new perspectives compared with schemes that rely on 
thermal effects only. \cite{Koenig:00,Wang:08,Kirilyuk:10}
While most of the coherent magnetization dynamics 
studied so far in extended semiconductors are dealing with the observation and the control of 
coherent spin precession, \cite{crooker:97,wang:09,Kapetanakis:09} 
the transfer of spin between the Mn and the electron or hole subsystems 
is usually considered to be an incoherent assimilation process that can adequately be described
by a Markovian rate. \cite{Krenn:85,Krenn:89,Cywi:07,Morandi:09} 
In contrast, for single quantum dots with a single embedded Mn atom
the spin transfer is a coherent process where spin is exchanged between
discrete levels by Rabi-type rotations.\cite{Reiter:09,Reiter:11,Reiter:12}
It will be shown in this paper that also in quantum wells and wires a
coherent  spin transfer between carriers and Mn atoms 
is possible where spins are exchanged back and forth between these subsystems.
Our analysis reveals a pivotal role of the energy-time uncertainty for enabling
the coherent spin-exchange in extended systems.

In this paper, we concentrate on the spin transfer between Mn atoms and electrons in the 
II-Mn-VI semiconductor ZnMnSe and disregard other mechanisms that may change the electronic spin.
This spin transfer process, of course, in general competes with many other processes that may affect 
the electronic spin dynamics such as the Dyakonov-Perel, Elliott-Yafet or Bir-Aronov-Pikus mechanism\cite{Dyakonov:08}  
which may lead to a spin relaxation.
In Ref.~\onlinecite{Camilleri:01}  it was shown that in the conduction band of II-Mn-VI semiconductors
the spin transfer can become the dominant intrinsic influence on the electron spin which justifies our
focus on this mechanism.

We shall demonstrate that the spin transfer may proceed qualitatively different from
what is predicted by a Markovian-rate equation even though the memory associated with this 
process is short. While a short memory time is a necessary precondition for the validity
of Markovian-rate theories we still find deviations from the rate behavior that can be 
traced back to the fact that the exchange interaction which is responsible for
the spin transfer between electrons and Mn atoms simultaneously leads to
a redistribution of electron energies that is lost in the Markov limit.
We have performed simulations of the pertinent dynamics for bulk, well, and wire samples
and find the most pronounced deviations from the Markov limit for quasi
one-dimensional wire systems. For wells the deviations are weaker but clearly visible while 
bulk sample are well described by the Markovian theory.
We are using for our calculations
a quantum kinetic theory of the correlated spin dynamics in diluted magnetic semiconductors (DMS) 
that we have recently developed which treats the exchange interaction between Mn atoms and carriers beyond the mean-field 
and virtual crystal approximations. \cite{Thurn:12}
While other approaches that explicitly account for correlations
have been derived in the framework of the Greens-function formalism, \cite{Kapetanakis:08a}
our approach is based on a microscopic density matrix  theory.

\section{\label{SP} Model}

In this paper, we apply the theory developed in 
Ref.~\onlinecite{Thurn:12} to the description of the
spin transfer dynamics in the conduction band of ZnMnSe.
We consider a single band model a Hamiltonian $H$ consisting of two parts:
 \begin{subequations}
\label{eq:1}
\begin{align}
\label{eq:1a}
 H &= H_0 + H_{sd},
\end{align}
the electronic band structure $H_0$ of the host semiconductor and the exchange 
interaction $H_{sd}$ between $s$-like conduction band electrons and $d$-like localized Mn orbitals.

The host semiconductor may be a bulk material, a quantum well or a quantum wire which,
when concentrating on the lowest subband, represent systems of effective
dimension $D=3,2,1$, respectively.
In all cases,  we account for two spin-degenerate conduction bands with a well defined spin quantum number of $s=\frac 1 2$.
In second quantization, $H_0$ then reads:
\begin{align}
\label{eq:1b}
H_0 &= \sum_{\sigma\bk} E_{\bk} c^\dagger_{\sigma \bk}c_{\sigma \bk},
\end{align}
where $\sigma \in \{\downarrow,\uparrow \}$ is the spin quantum number, $\bk$ denotes the $D$-dimensional wave-vector and 
$E_\bk$ the corresponding carrier energy. 

The Mn atoms have half-filled $d$-shells with vanishing angular momenta and can be described as localized $S=\frac{5}2$ {}
spins in good approximation. The exchange interaction in contact form between these spins and the conduction band 
electrons is given by the Kondo-like Hamiltonian:
\begin{align}
\label{eq:1c}
H_{sd} &=  \frac{J_{sd}}V \sum_{Inn' \atop \sigma\sigma'{\bk}{\bk}'} {\bS}_{nn'}\cdot  
{\bs}_{\sigma\sigma'}  e^{i({\bk}' - {\bk}){\bR}_I} c^\dagger_{\sigma{\bk}}c_{\sigma'{\bk}'} \hat P_{nn'}^I,
\end{align}
\end{subequations}
where $J_{sd}$ is the exchange constant, $V$ the sample Volume, ${\bS}_{nn'}$ the vector of Mn spin 
matrices with $n \in \{-\frac 5 2, \dots, \frac 5 2\}$. 
${\bs}_{\sigma\sigma'} $ denotes the vector of spin Pauli matrices
and $\bR_I$ stands for the position vector of the $I$-th Mn atom.
The operator
\begin{align}
\label{eq:2}
\hat P_{nn'}^I :=  |I,n\rangle \langle I,n'|
\end{align}
represents the Mn spin degrees of freedom, where $|I,n\rangle$ denotes the $n$-th eigenstate of the z-component of 
the $I$-th Mn spin. In real samples the Mn dopants are randomly placed all over the sample. The Hamiltonian
$H_{sd}$ is formulated for a given specific configuration.

Starting from the Hamiltonian in Eq.~\eqref{eq:1}, we have derived  a closed set of 
quantum kinetic equations of motion [Eqs.~(16)-(17) in Ref.~\onlinecite{Thurn:12}] which describe correlated Mn and electron dynamics
in an on average spatially homogeneous system.\cite{Thurn:12}  The averaging involves a quantum mechanical average as well as
an average over the random  Mn positions which are assumed to  be homogeneously distributed in space.

In order to focus on the spin transfer dynamics we consider a situation where initially all electron spins are aligned
along the $z$ axis and the Mn spin are equally distributed over the six spin states leading to a zero Mn magnetization.
 With these initial conditions no precession will occur as the electron spin vector
will stay at all times parallel to the $z$-axis and to the  Mn spin vector 
that will acquire finite values in the course of time due to the spin transfer.
As a consequence, it turns out that only the following
subset of the dynamic variables of Eqs.~(16)-(17) in Ref.~\onlinecite{Thurn:12} has
non-zero values: 
\begin{subequations}
\label{eq:3}
\begin{align}
  \label{eq:3a}
&\M{}{}  := \langle \hat P_{nn}^I\rangle, \\[1ex]
  \label{eq:3b}
&\C{}{}{}  := \langle c^{\dag}_{\sigma\bk}c_{\sigma\bk}\rangle, \\[1ex]
  \label{eq:3c}
  &\Ck{}{}{_{1}}{_{2}}:=V \delta \langle c^{\dag}_{\sigma\bk_{1}}c_{\sigma\bk_{2}} e^{i(\bk_{2} - \bk_{1})\bR_{I}}\rangle
% \nn
% & = V \langle c^{\dag}_{\sigma\bk_{1}}c_{\sigma\bk_{2}} e^{i(\bk_{2} - \bk_{1})\bR_{I}}\rangle - 
% \delta_{\bk_1\bk_2} \langle c^{\dag}_{\sigma\bk_{1}}c_{\sigma\bk_{2}}\rangle
, \\[1ex]
  \label{eq:3d}
  &\K{_{1}}{_{2}}{_{1}}{_{2}}{_{1}}{_{2}} := V \delta \langle c^{\dag}_{\sigma_{1}\bk_{1}}c_{\sigma_{2}\bk_{2}} \hat P_{n_{1}n_{2}}^I e^{i(\bk_{2} 
- \bk_{1})\bR_{I}}\rangle,
\end{align}
\end{subequations}
where $\langle \dots \rangle$ stands for both the quantum mechanical average with respect to the statistical operator 
and the disorder average with respect to the random spatial distribution of the Mn atoms.  
The correlation functions $\delta \langle c^{\dag}_{\sigma\bk_{1}}c_{\sigma\bk_{2}} e^{i(\bk_{2} - \bk_{1})\bR_{I}}\rangle$
and
$\delta \langle c^{\dag}_{\sigma_{1}\bk_{1}}c_{\sigma_{2}\bk_{2}} \hat P_{n_{1}n_{2}}^I e^{i(\bk_{2} - \bk_{1})\bR_{I}}\rangle$
occurring in Eqs.~\eqref{eq:3c} and \eqref{eq:3d}
are related to the corresponding expectation values  $\langle c^{\dag}_{\sigma\bk_{1}}c_{\sigma\bk_{2}} e^{i(\bk_{2} - \bk_{1})\bR_{I}}\rangle$
and 
$\langle c^{\dag}_{\sigma_{1}\bk_{1}}c_{\sigma_{2}\bk_{2}} \hat P_{n_{1}n_{2}}^I e^{i(\bk_{2} - \bk_{1})\bR_{I}}\rangle$
by subtracting from the latter all possible factorizations into factors involving expectation values of fewer operators. 
The lengthy explicit definitions of these correlation functions
can be found in Eq.~(13) of Ref.~\onlinecite{Thurn:12}.

$\M{}{}$ represents the average occupation of the $n$-th Mn eigenstate and 
$\C{}{}{}$ the occupation of the Bloch state with wave vector $\bk$ and spin $\sigma$. 
$\Ck{}{}{_{1}}{_{2}}$ describes  a correlation between electrons and Mn positions $\bR_I$ 
(averaged over the random positions of the Mn atoms) which is due to the 
scattering of the electrons with the localized Mn atoms.
This correlation function would also arise if the interaction between electrons and Mn spins would
be given by a spin independent localized potential at random Mn positions instead of the exchange
interaction Eq.~\eqref{eq:1c}.
$\K{_{1}}{_{2}}{_{1}}{_{2}}{_{1}}{_{2}}$ describes correlations between electrons, Mn spins and their respective positions. 
In this function also spin correlations between the Mn and the electron subsystems are represented that arise because
the exchange interaction in addition to providing a localized scattering channel acts on the spin degree of freedom.

Using this reduced set of variables the pertinent equations of motion can be derived for bulk, well and wire systems
along the lines detailed in  Ref.~\onlinecite{Thurn:12} resulting for the $D$ dimensional case in: 

\begin{widetext}
\begin{subequations}
\label{eq:5}
\begin{align}
\label{eq:5a}
- i \hbar \dt \M{_1}{_1} &= 
\frac {J_{sd}L^{2D}} {V^2(2\pi)^{2D}} \iint\limits_{BZ}  \sum_{n\sigma\sigma'} \bs_{\sigma  \sigma'}  \cdot  
\Big(\bS_{nn_1}  \K{}{'}{}{_1}{}{'}   
-  \bS_{n_2n}  \K{}{'}{_1}{}{}{'}\Big)    \dkd\dkd', \allowdisplaybreaks\\[2ex]
%%%%%%%%%%%%%%%%%%%%%%%%%%%%%%%%%%%%%%%%%%%%%%%%%%%%%%%%%%%%%%%
%%%%%%%%%%%%%%%%%%%%%%%%%%%%%%%%%%%%%%%%%%%%%%%%%%%%%%%%%%%%%%%
\label{eq:5b}
-i\hbar \dt  \C{_1}{_1}{_1} &=   n_{\text{Mn}} \frac { J_{sd} L^D} {V(2\pi)^D} \int\limits_{BZ} \sum_{n} \Bigg\{ 
{\bS}_{nn} \cdot   {\bs}_{ \sigma_1 \sigma_1}
 \M{}{} \Big(   \Ck{_1}{_1}{}{_1}
-  \Ck{_1}{_1}{_1}{}  \Big) 
+ 
  \sum_{n'\sigma} {\bS}_{nn'} \cdot   \Big(  {\bs}_{ \sigma \sigma_1} \K{}{_1}{}{'}{}{_1} 
-  {\bs}_{ \sigma_1 \sigma} \K{_1}{}{}{'}{_1}{}  \Big)     \Bigg\}\dkd  , \allowdisplaybreaks \\[2ex]
%%%%%%%%%%%%%%%%%%%%%%%%%%%%%%%%%%%%%%%%%%%%%%%%%%%%%%%%%%%%%%%
%%%%%%%%%%%%%%%%%%%%%%%%%%%%%%%%%%%%%%%%%%%%%%%%%%%%%%%%%%%%%%%
%
%
%
\label{eq:5c}
\big( -i\hbar \dt  + & E_{\bk_{2}}-  E_{\bk_{1}} \big) \Ck{_1}{_1}{_1}{_2}=  
 J_{sd}  \sum_{n}  \Bigg\{ {\bS}_{nn} \cdot {\bs}_{\sigma_1\sigma_1} 
 \M{}{} \bigg[
  \C{}{_1}{_2} - \C{_1}{_1}{_1} +  \frac{L^D}{V(2\pi)^D} \int\limits_{BZ}  \Big(\Ck{_1}{_1}{}{_2} -  \Ck{_1}{_1}{_1}{}\Big)\dkd \bigg]
\nn 
& +
\sum_{n' \sigma}  \frac{L^D}{V(2\pi)^D}\int\limits_{BZ} {\bS}_{nn'} \cdot \Big(  {\bs}_{\sigma\sigma_1}  
 \K{}{_1}{}{'}{}{_2} 
 -  {\bs}_{\sigma_1\sigma}  
 \K{_1}{}{}{'}{_1}{}  \Big) 
\dkd 
 \Bigg\}, 
\allowdisplaybreaks \\[2ex]
%%%%%%%%%%%%%%%%%%%%%%%%%%%%%%%%%%%%%%%%%%%%%%%%%%%%%%%%%%%%%%%
%%%%%%%%%%%%%%%%%%%%%%%%%%%%%%%%%%%%%%%%%%%%%%%%%%%%%%%%%%%%%%%
%
%
%
\label{eq:5d}
\Big(-i\hbar \dt  + &E_{\bk_2} - E_{\bk_1} \Big)  \K{_1}{_2}{_1}{_2}{_1}{_2}  = 
J_{sd}  
 {\bS}_{n_2n_1} \cdot  {\bs}_{ \sigma_2 \sigma_{1}} \Big(    
\C{_2}{_2}{_2} \M{_2}{_2} - \C{_1}{_1}{_1} \M{_1}{_1}
\Big) 
+
 Q_{\K{_1}{_2}{_1}{_2}{_1}{_2}},
%%%
\end{align}
\end{subequations}
\end{widetext}
where the source $ Q_{\K{_1}{_2}{_1}{_2}{_1}{_2}}$ on the right hand side of Eq.~\eqref{eq:5d} is given 
explicitly in the appendix. 
The integrals are to be taken over the first Brillouin Zone ($BZ$). $L$ denotes the length of the system in the unconfined directions, 
$V = L^3$ for bulk systems, $V = L^2 d$ for quantum wells of the thickness $d$ and $V = L A$ for quantum wires with a cross sectional area $A$.
For the numerical implementation of the above equations it is advantageous to note that
in the  case considered here,  $\bar K$ is only non-zero if the indices $\sigma$ and $n$ fulfill either
\begin{subequations}
\label{eq:4}
\begin{align}
\label{eq:4a}
\sigma_1 &= \sigma_2, \quad n_1 = n_2 \qquad \text{or}, \\
\label{eq:4b}
\sigma_1 &= \sigma_2 \pm 1, \quad n_1 = n_2 \mp 1, \quad n_1 + \sigma_1 =  n_2 + \sigma_2.
\end{align}
\end{subequations}

\section{General properties of the equations of motion}
\subsection{Relative importance of different correlations}

In this subsection we shall demonstrate that for the spin transfer processes to be discussed in this paper
only a much smaller subset of terms essentially contributes. The resulting reduction in complexity is 
not only advantageous for the numerics, it will also enable us to analyze more conclusively the
pertinent features of the resulting dynamics. 

The key observation for identifying the most important contributions is that
for an initially uncorrelated system in which the average Mn spin is zero and the 
electrons are completely spin polarized the onset of the spin transfer is mediated exclusively 
by the first source term in Eq.~\eqref{eq:5d} for the correlation function $\bar K$
\begin{align}
\label{eq:7}
 J_{sd} {\bS}_{n_2n_1} \cdot  {\bs}_{ \sigma_2 \sigma_{1}} \Big(    
\C{_2}{_2}{_2} \M{_2}{_2} - \C{_1}{_1}{_1} \M{_1}{_1}
\Big),
\end{align}
as has been shown in Ref.~\onlinecite{Thurn:12}. 
This implies that this term plays a pivotal role for the spin transfer dynamics that is
the target of the present paper. The resulting  $\bar K$ directly drives via Eqs.~\eqref{eq:5a}
and \eqref{eq:5b} the main observables of interest $M$ and $C$.
The rather involved additional sources 
$Q_{\bar K}$  (cf.~the appendix) in Eq.~\eqref{eq:5d} as well as the disorder related
correlations $\bar C$ build up subsequently in a second step and are of higher order in the 
coupling constant $J_{sd}$.\footnote{Note, that for a ferromagnetic material, where initially the 
total Mn magnetization has a finite value, additional terms of order $\mathcal{O}(J_{sd})$
arise.}
Based on this observation it is suggestive to assume that
$Q_{\bar K}$ and $\bar C$ might be of less importance also when the dynamics is followed over
a longer time scale, although the correlation functions $Q_{\bar K}$ and $\bar C$ appear on the same
level of the correlation expansion as the remaining contributions.
In order to test this assumption, 
we have performed numerical simulations where we compare results of the full set of equations 
Eqs.~\eqref{eq:5} with calculations where $Q_{\bar K}$ and $\bar C$ have been set  to zero.

To be specific, we have numerically solved the initial value problem of initially spin polarized 
electrons with a Gaussian distribution in energy space according to
\begin{subequations}
\label{eq:8}
\begin{align}
\label{eq:8a}
\C{}{}{}|_{t_0} = \delta_{\sigma\uparrow} \exp\Big[-\frac{(E_\bk - E_0)^2}{2\Delta^2}\Big],
\end{align}
and Mn spins where initially all possible $z$ components have equal probabilities, i.e.
\begin{align}
\label{eq:8b}
\M{}{}|_{t_0} = \frac 1 6.
\end{align}
Finally, all correlations are initially set to zero
\begin{align}
\label{eq:8c}
\Ck{}{}{}{}|_{t_0} = \K{}{}{}{}{}{}|_{t_0} = 0.
\end{align}
\end{subequations}
These initial conditions are meant to mimic the situation immediately after a fast optical excitation
of an initially unmagnetized DMS.

\begin{figure*}[t]
\begin{centering}
\includegraphics[width=\textwidth]{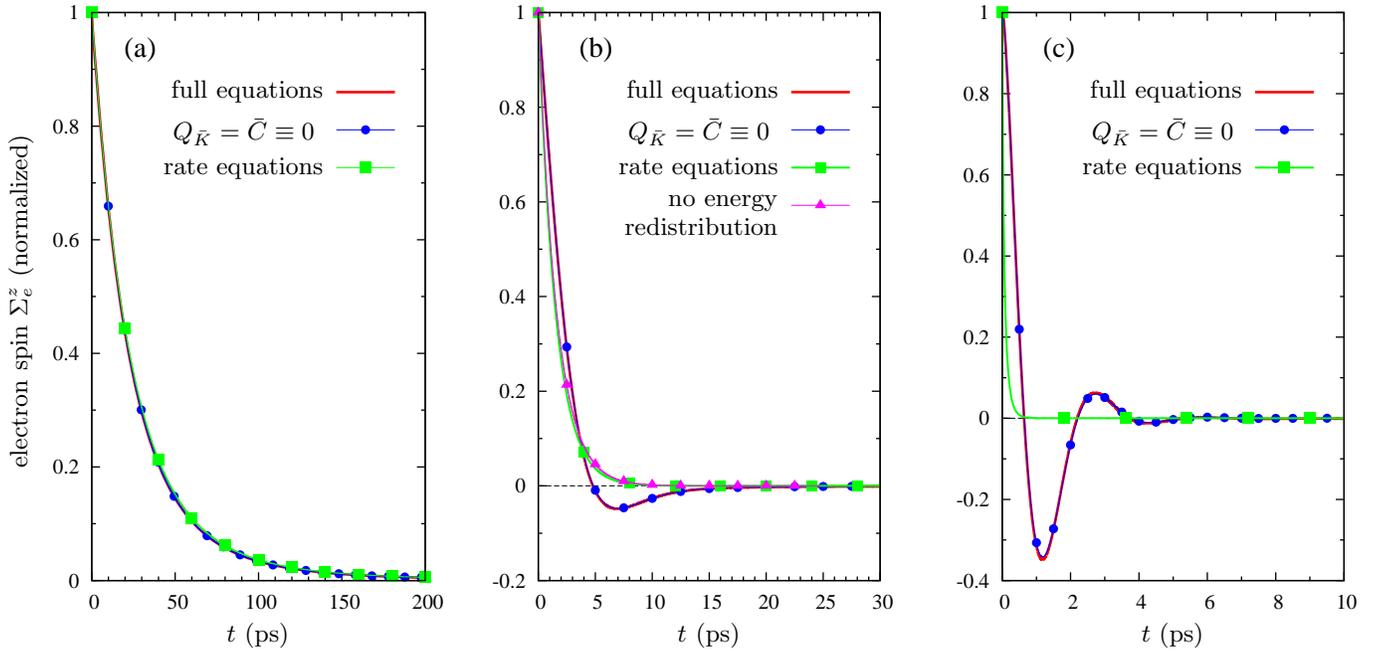}
\par\end{centering}
 \caption{(color online) Time evolution of the total electron spin $\Sigma_e^z$ in Zn$_{0.93}$Mn$_{0.07}$Se 
for (a) a bulk system; (b) a quantum well; (c) a quantum wire. The solid red lines have been calculated with the full 
quantum kinetic Eqs.~\eqref{eq:5}, the blue lines with bullets have been obtained by neglecting $Q_{\bar K}$ and 
$\bar C$ in Eqs.~\eqref{eq:5}, and the green lines with squares represent the results of
the Markovian rate Eqs.~\eqref{eq:16}. The pink curve with triangles in (b) has been obtained from Eqs.~\eqref{eq:21},
where the energetic redistribution of electrons is neglected.
}
\label{fig:1} 
\end{figure*} 

We have performed simulations for 
a wide range of parameters typical for DMS.  
In particular, we have varied the exchange coupling $J_{sd}$ 
in the range 5-100 meV$\text{nm}^{3}$ and the Mn concentration $x$
in the range of 1-10~\% and used 
an effective electron mass of $m_e = 0.21 m_0$ with
$m_0$ being the free electron mass, 
which is typical for Zn$_{1-x}$Mn$_x$Se.
The parameters $\Delta$ and $E_{0}$
of the initial Gaussian distribution in Eq.~\eqref{eq:8a}
have been taken to be $\Delta = 0.4$~meV and $E_0 = 0$~meV,
respectively. The above initial value problem has been solved
for bulk, a well with width $d = 4$ nm and a wire with a cross-section of 
area $A = 16$~nm$^2$.
It turns out that the results are qualitatively similar in the whole parameter range studied.
Typical results that correspond to Zn$_{0.93}$Mn$_{0.07}$Se where
 $J_{sd}=12$~meV$\text{nm}^{3}$ are shown in Fig.~\ref{fig:1}, where
the total electron spin 
\begin{align}
\label{eq:9}
{\bm{\Sigma}}_e = \sum_{\bk\sigma} \bs_{\sigma\sigma} \C{}{}{}
\end{align}
is plotted as a function of time for 
(a) a bulk semiconductor, (b) a quantum well and (c) a quantum wire.
The solid red lines represent results of the full set of equations while the
blue thin lines with bullets are obtained by neglecting  $Q_{\bar K}$ and $\bar C$.
The green lines with squares are obtained from
Markovian rate equations and will be discussed later.
At this point we  only want to note that the results of the former two calculations
quantitatively are almost indistinguishable for all conditions studied.
For a detailed discussion of the physical implications of the curves in Fig.~\ref{fig:1}
it turns out to be useful to exploit the fact that $Q_{\bar K}$ and $\bar C$
can be safely neglected in our case for a reformulation of the
remaining equations.

\subsection{Integral representation and memory kernel}
\label{IRAMK}

When $Q_{\bar K}$ and $\bar C$ are discarded the right hand side of the equation of motion \eqref{eq:5d} for
$\bar K$  is independent of $\bar K$ and can easily be integrated resulting in:
\begin{align}
\label{eq:10}
&\K{_1}{_2}{_1}{_2}{_1}{_2} = J_{sd} \int_{t_0}^t \frac i \hbar
e^{ \frac{i}{\hbar}(E_{\bk_1} - E_{\bk_2})(t - t')} {\bS}_{n_2n_1} \cdot  {\bs}_{ \sigma_2 \sigma_{1}} \nn
& \qquad  \times \Big(    
\C{_2}{_2}{_2}(t') \M{_2}{_2}(t') - \C{_1}{_1}{_1}(t') \M{_1}{_1}(t')\Big)
 dt'. 
\end{align}
Here we have again assumed that $\K{_1}{_2}{_1}{_2}{_1}{_2}(t_0) = 0$.
Inserting Eq.~\eqref{eq:10} in 
the equations of motion \eqref{eq:5a} and \eqref{eq:5b} for $C$ and $M$
formally eliminates $\bar K$ from these equations leading to a closed set 
of equations of motion involving only $C$ and $M$ as dynamical variables:

\begin{widetext}
\begin{subequations}
\label{eq:11}
\begin{align}
\label{eq:11a}
& \dt \M{_1}{_1} 
=
\frac {2 J_{sd}^2L^{2D}} { \hbar^{2} V^2(2\pi)^{2D}}  
\sum_{n\sigma\sigma'}  \big( \bS_{nn_1} \cdot  \bs_{\sigma\sigma'}  \big)^2    
%\nn & \times\! \!
\int\limits_{BZ}\!\int\limits_{t_0}^t 
\!G_\bk^D(t\!-\!t')
\Big( 
\C{}{}{}(t')\M{}{}(t') 
-  
\C{'}{'}{}(t')\M{_1}{_1}(t')  
 \Big)dt' \dkd, 
\allowdisplaybreaks\\[1.5ex]
\label{eq:11b}
&\dt  \C{_1}{_1}{_1}
\!=
\frac {2J_{sd}^2  n_{Mn} L^D} {\hbar^{2} V (2\pi)^D}\!\! \int\limits_{BZ}\!\int\limits_{t_0}^t \sum_{nn'\sigma} 
\! \big(\bS_{nn'} \cdot \bs_{\sigma\sigma_1} \big)^2 
% \nn &\quad\times
\cos\! \Big( \frac{E_{\bk}\! -\! E_{\bk_1}}{\hbar}(t\! -\! t')\! \Big) 
 \Big(   
\C{}{}{}(t')\M{}{}(t')\! 
\!-\!   
\C{_1}{_1}{_1}\!(t') \M{'}{'}(t')
\Big) 
% \nn &\quad \times   
dt'\dkd,
\end{align}
\end{subequations}
\end{widetext}
where the $\bk$-dependent memory function $G_\bk^D(\tau)$ is defined  as:   
\begin{align}
\label{eq:12}
G_\bk^D(\tau) &:= \int\limits_{BZ} \cos\Big(\frac 1 \hbar\big(E_{\bk'}-E_{\bk}\big)\tau\Big) \dkd'.
% \nn
% & =  \Omega_D\int_0^{K} \cos\Big(\frac{\hbar(k'^2 - k^2)}{2m_e}\tau\Big) k'^{D-1} dk'
\end{align}

\begin{figure}[tbph]
\begin{centering} 
\resizebox{\columnwidth}{!}{\includegraphics{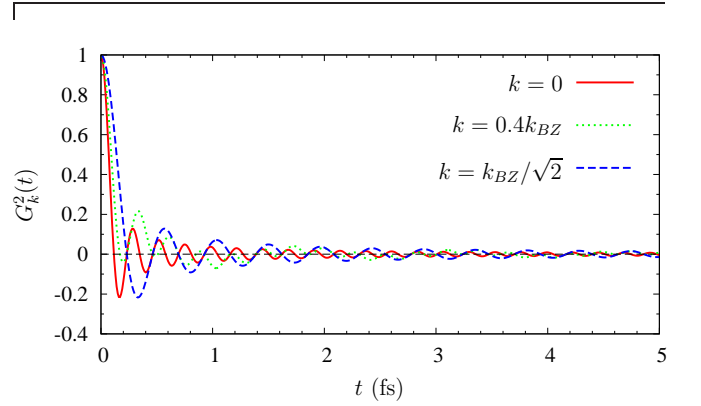}}
\par\end{centering} 
 \caption{(color online) Spin memory function $G_k^2(\tau)$ of a ZnMnSe quantum well for different values of $k$.}
\label{fig:2}
\end{figure}

If one is  interested in the total electron spin ${\bm\Sigma}^{e}$ in the system Eq.~\eqref{eq:11b} has to be summed over $\bk$.
After this summation the same memory function $G_\bk^D(\tau)$ as in Eq.~\eqref{eq:11a}
appears also in the equation for ${\bm\Sigma}^{e}$ demonstrating that $G_\bk^D(\tau)$ determines the
memory of the spin transfer process.
Formulating the equations of motion using a memory kernel 
has the advantage that we can now analyze the corresponding memory depth. 
In the case of a quantum well this can be made even more explicit as
for parabolic bands (i.e. $E_\bk = \frac{(\hbar\bk)^2}{2m_e}$)
and approximating the Brillouin zone as a sphere with radius $k_{BZ} = |\bk_{BZ}|$
the integral in  Eq.~\eqref{eq:12} can be 
evaluated analytically resulting in:
\begin{align}
\label{eq:14}
G_k^2(\tau) 
&= \frac{4\pi m_e}{\hbar\tau} \sin\Big( \frac{k_{BZ}^2}{4m_e}\hbar\tau\Big) \cos\Big(\frac{k_{BZ}^2 - 2k^2}{4m_e}\hbar\tau \Big).
\end{align}
Fig.~\ref{fig:2} shows the memory kernel $G_k^2(\tau)$ as a function of $\tau$ using
the ZnMnSe effective mass $m_e = 0.21 m_0$ for 
three representative values of $k$ [$k=0$  ($k=k_{BZ}/\sqrt{2}$)  
correspond to the $k$ values where the first zero of $G_k^2(\tau)$
comes at the earliest (latest) time while $k=0.4 k_{BZ}$ is an intermediate
value]. In all cases $G_k^2(\tau)$ decays and oscillates rapidly on a sub-femtosecond timescale and 
depends only weakly on $k$. These general statements also hold for bulk and quantum wire systems, 
where we have calculated $G_\bk^D(\tau)$ numerically.

As can be seen in Fig.\ref{fig:1}, the timescale of the exchange induced spin dynamics in typical DMS is of the order of
several picoseconds, i.e. over $10^3$ times larger than the memory depth of the system. One could, thus, suspect that it is
justified to treat the spin transfer dynamics in the Markovian limit.

\subsection{Markovian limit}

In this subsection we shall derive the Markovian limit of the spin transfer described by our model
and compare the results with our quantum kinetic theory represented by Eqs.~\eqref{eq:5} that
here give essentially the same results as the non-Markovian Eqs.~\eqref{eq:11} (cf.~Fig.~\ref{fig:1}).

The Markov limit corresponds to neglecting the retardations
in Eq.~\eqref{eq:10} of the functions $M$ and $C$. 
Formally, this is done by evaluating $M$ and $C$ at time $t$ 
instead of $t'$ which allows to take these functions out of the 
integral over the past. Finally, the limit $t_{0}\to-\infty$ is taken which
eliminates the initial time. The remaining integral yields $\delta$-distributions
and principal value integrals. The latter do however drop out when
inserted in the equations of motion for $M$ and $C$. The principal parts also
vanish in the limit $t\to \infty$ which is the limit usually discussed when deriving
Fermi's golden rule.  For a parabolic bandstructure this procedure yields  
in Eq.~\eqref{eq:10}: 
\begin{align}
\label{eq:15}
 &\K{_1}{_2}{_1}{_2}{_1}{_2} \approx J_{sd}  \bS_{n_2n_1}\bs_{\sigma_2\sigma_1}  \Big(   \C{_2}{_2}{_2}\M{_2}{_2}
- \C{_1}{_1}{_1}\M{_1}{_1} \Big) \nn & \qquad \times i\pi\delta\Big(\frac{\hbar^2}{2m_e}(\bk_1^2 - \bk_2^2)\Big).
\end{align}
It should be noted that going over to the Markov limit introduces the energy conserving delta function
in Eq.~\eqref{eq:15} which for parabolic bands makes $\bar K$ diagonal with respect to the
absolute values of $\bk_{1}$ and $\bk_{2}$.

With $\bar K$ approximated by Eq.~\eqref{eq:15}
the equations of motion~\eqref{eq:11a} and \eqref{eq:11b} for $M$ and $C$ then become:
\begin{subequations}
\label{eq:16}
\begin{align}
\label{eq:16a}
& \dt \M{_1}{_1} 
= \frac { J_{sd}^2 m_e \Omega_D  L^{2D}} {\hbar^3 V^2 (2\pi)^{2D-1}}  
\sum_{n\sigma\sigma'} \big( \bS_{nn_1} \cdot  \bs_{\sigma\sigma'}  \big)^2  \nn
& \qquad \times
\int\limits_{BZ} \Big(  C^\sigma_{\sigma k}\M{}{} -  C^{\sigma'}_{\sigma' k}\M{_1}{_1}
 \Big)   k^{D-2} dk,\allowdisplaybreaks\\[1.5ex]
\label{eq:16b}
& \dt  C^{\sigma_1}_{\sigma_1 k_1}
= \frac {J_{sd}^2  m_e  \Omega_D L^D} {\hbar^3 V(2\pi)^{D-1}} n_{Mn}  \sum_{nn'\sigma} 
\big(\bS_{nn'}\cdot\bs_{\sigma\sigma_1} \big)^2 \nn & \qquad \times
\Big(  C^{\sigma}_{\sigma k_1} \M{}{} -  C^{\sigma_1}_{\sigma_1 k_1}\M{'}{'}\Big)  k_1^{D-2},
\end{align}
where $\Omega_3 = 4\pi$, $\Omega_2 = 2\pi$ and $\Omega_1 = 2$.
In the above equation, we have assumed the bandstructure to be parabolic and 
averaged the electronic variables with respect to the angle of their $\bk$-vectors: 
$$C^{\sigma}_{\sigma k} = \frac {\int \C{}{}{} d\Omega_D} {\Omega_D}.$$ For such a bandstructure, 
this averaging procedure directly leads to 
a closed set of equations of motion without requiring any further approximations. 
Since the number of Mn atoms in an DMS is usually several orders of
magnitude larger than the number of optically generated spin polarized electrons, the Mn variables $\M{}{}$ are almost 
constant.\footnote{Our numerical simulations show that for parameter range discussed in this paper, it is always 
viable to set $M$ constant. This holds for both, the quantum kinetic and the Markovian calculations.}
In the paramagnetic case with zero magnetic field,  $\M{}{}(t)$ is approximately $\frac 1 6$. 
\end{subequations}
Then, it follows from Eq.~\eqref{eq:16b} that the spin of an electron with momentum $\hbar k_{1}$ is given by 
$s^z_{k_1} = s^z_{\uparrow\uparrow}C^\uparrow_{\uparrow k_1} 
- s^z_{\downarrow\downarrow}C^\downarrow_{\downarrow k_1}$  
and simply decays exponentially:
\begin{align}
\label{eq:17}
 s^z_{k_1}(t)
=
  s^z_{k_1}(t=0) e^{-\gamma_{k_1} t}
\end{align}
with the rate 
\begin{align}
\label{eq:18}
 \gamma_{k_1} = \Omega_D \frac {35\pi J_{sd}^2  m_e L^D} {6\hbar^3 V(2\pi)^D} n_{Mn}  k_1^{D-2}.
\end{align}
This rate, which we have obtained as a limiting case of our full quantum kinetic equations~\eqref{eq:5}, is already 
known since the 1970s where it has been derived from a golden rule analysis for the bulk case.\cite{Kossut:75} For a quantum
well the rate in Eq.~\eqref{eq:18} has been deduced in, e.g., Ref.~\onlinecite{Jiang:09}.

Now, the quantum kinetic Eqs.~\eqref{eq:5} can be used to check the validity of the rate 
Eqs.~\eqref{eq:16} for the description of spin transfer dynamics in different DMS systems. 
Fig.~\ref{fig:1}  shows
the electron spin dynamics in Zn$_{1-x}$Mn$_x$Se systems of different dimensionality
and compares quantum kinetic results (thick solid lines) with the Markov limit
represented by the rate Eqs.~\eqref{eq:16} (green dashes lines with squares).
As can be seen:  (a)  the rate equations are in excellent agreement
with quantum kinetic equations for bulk semiconductors, 
(b) for a quantum well the full equations predict a clear overshoot
of the electron spin which is absent in the Markovian limit and 
(c)  according to the full equations, the electron spin in a quantum wire can even show oscillations 
that are also not expected from rate equations.

\begin{figure}[htbp]
\begin{centering}
\includegraphics[width=\columnwidth]{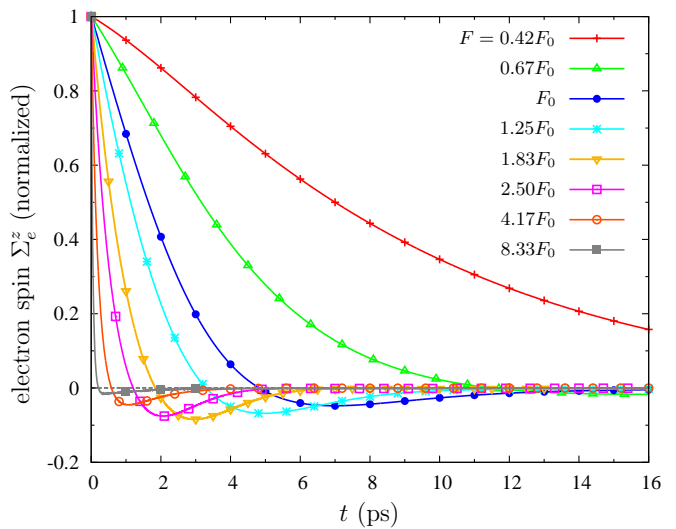}
 \caption{(color online) Time evolution of the total electron spin $\Sigma_e^z$ for different values of the rescaled 
electron Mn coupling $F$ [cf. Eq.~\eqref{eq:19}] at fixed mass ratio $m_{0}/m_{e}$. $F_{0} $ corresponds to
the parameters used in Fig.~\ref{fig:1} (b).
}
\label{fig:3}
\par\end{centering} 
\end{figure}

Before analyzing in detail the origin of these non-Markovian features,  
we shall  shortly examine the impact of material parameters. 
If we again approximate $M$ as constant and rescale the time $t$ with $m_0 /m_e $, it can be 
seen from Eq.~\eqref{eq:11b} that material parameters enter the electron spin dynamics only as the prefactor 
\begin{align}
\label{eq:19}
F =  m_e J_{sd}^2 n_{Mn} L^D V^{-1} 
\end{align}
of the source terms on the right hand side. 
Here, we  illustrate the influence of material parameters exemplarily by
simulating the initial value problem from Eqs.~\eqref{eq:8} for  quantum wells
with different prefactors $F= m_e J_{sd}^2 n_{Mn}  d^{-1}$ but fixed mass ratio $m_0 /m_e $. 
Displayed in Fig.~\ref{fig:3} is the resulting total electron spin as a function of the unscaled time $t$.
It can be clearly seen that the amplitude of the overshoot non-monotonically depends on $F$, while the time it takes 
until the maximal overshoot is reached decreases with $F$.  
For small values of $F < 0.5 F_0$, practically no visible overshoot takes place and the results are in good agreement with the
rate equations \eqref{eq:16}, for $F \approx 1.8 F_0$ the overshoot 
amplitude reaches its maximum before it decreases with increasing $F$. 
A similar trend is found for quantum wires, where the strength of the oscillations
depends non-monotonically on $F$ (not shown).

The spin overshoot and oscillations are clear signatures
of coherent dynamics which cannot be described on the level of rate equations. 
It should be recalled that the oscillations
are not related to a precession of the spins which is not possible for the configuration considered here
as the Mn and electron spins are aligned parallel. Instead they represent a coherent exchange of spin
between the electronic and the Mn subsystems.
It is also worth noting that the rate equations 
fail in the two- and one-dimensional case even despite the femtosecond spin 
memory depth revealed by Eq.~\eqref{eq:12} and seen in Fig.~\ref{fig:2} for $D=2$.

\section{Origin of the spin overshoot}
As a short memory time is commonly believed to guarantee the validity of the Markov approximation the question
arises, why the rate equations fail to describe the dynamics in the two- and one-dimensional systems  although 
this condition is fulfilled. It turns out that the corresponding analysis yields qualitatively the same
answers for wells and wires. Here we shall discuss explicitly only the case of  quantum wells as these systems 
are more widespread while Mn doped quantum wires are still a very novel field of research. \cite{Zaleszczyk:08}

The key to understand the origin of the observed non-Markovian behavior is to analyze not only
the total electron spin ${\bm{\Sigma}}_e$, which is obtained 
as a summation over $\bk$ space [cf.~Eq.~\eqref{eq:9}]
but to follow the time evolution of the
electron occupation  and the electron spin over the kinetic energies $E_{\bk}$ of the electrons.

\begin{figure}[hptb]
\begin{centering}
\begin{minipage}{\columnwidth}
\includegraphics[width=1\textwidth]{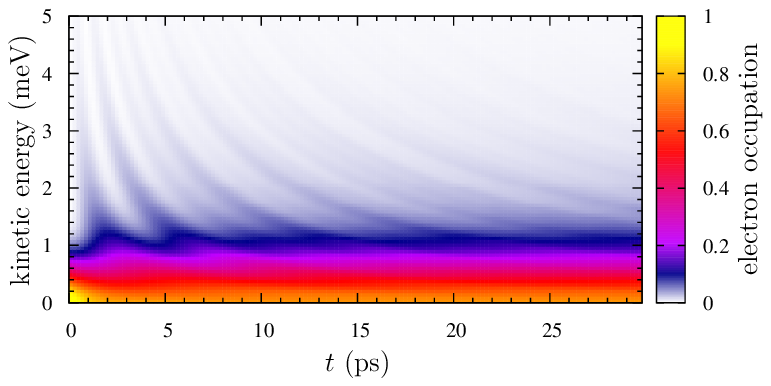}
\vspace{-0.4cm}
 \caption{(color online) Time evolution of the electron distribution in energy space. }
\label{fig:4}
\vspace{7mm}
\end{minipage}
\begin{minipage}{\columnwidth}
\includegraphics[width=1\textwidth]{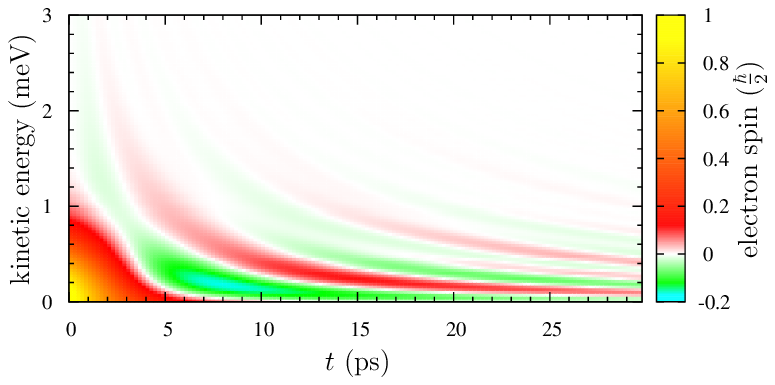}
\vspace{-0.4cm}
 \caption{(color online) Time evolution of the electron spin distribution in energy space. The electron spin is given in 
multiples of $\frac \hbar 2$. }
\label{fig:5}
\end{minipage}
\par\end{centering} 
\end{figure}

 Fig.~\ref{fig:4} shows the electron distribution as a function of  $E_\bk$
and the time $t$ for the parameters used for Fig.~\ref{fig:1}~(b). 
Initially, the electrons are Gaussian distributed. They are subsequently scattered towards 
higher energies, followed by a reflux. 
These  tails occur repeatedly before a quasi static electron distribution 
is reached for $t > 25$~ps. Eventually 19 \% of the electrons in the system
have been redistributed within the Brillouin zone. 

This  energetic redistribution is due to the scattering of electrons with the
spatially localized potentials of the Mn atoms. As discussed 
in Ref.~\onlinecite{Thurn:12} 
even in the limit of a spatially homogeneous Mn distribution which result in an on average 
spatially homogeneous system $\bk$-space scattering takes place as a result of the localization
of the Mn scattering centers. This scattering has been shown to be accompanied by a redistribution
over electronic energies \cite{Thurn:12} which in a quantum kinetic description is allowed due 
to the energy-time uncertainty.
We note that, according to the rate equation~\eqref{eq:16b}, no such electron redistribution 
is predicted since in the Markovian limit the kinetic energy is conserved in the scattering process
[cf.~Eq.~\eqref{eq:15}].

The electron spin distribution in energy space that 
corresponds to the electron distribution in Fig.~\ref{fig:4}
is depicted in Fig.~\ref{fig:5}.
It does not simply decay exponentially as would be expected in the Markovian limit 
according to Eq.~\eqref{eq:17}.  
Instead, the spin distribution forms several tails  in the course of time which correspond one-to-one
with similar tails of the electron occupations in Fig.~\ref{fig:4}. 
Each tail can be associated  either with a positive or 
negative electron spin. The sign of the spin of consecutive tails alternates, with the spin of the first tail being 
opposite to the positive initial spin. 
This can be explained by the nature of the interaction $H_{sd}$ of the electrons with the scattering 
centers, i.e. with the localized Mn spins. 
According to Eq.~\eqref{eq:1c}, the scattering of the itinerant electrons in $\bk$ space 
can take place with or without a flip of their spins due to the spin exchange with the Mn atom.
Analyzing the structure of the involved spin matrices reveals that 
processes with spin flip occur more often than those without spin flip.
Hence, starting initially with a positive spin the first tail resulting from the scattering 
is predominantly negative. 
The second tail contains mainly electrons that have been rescattered from the first tail to higher energies
and consequently their spin is positive.

\begin{figure*}[tp]
\begin{centering}
                \includegraphics[width=\textwidth]{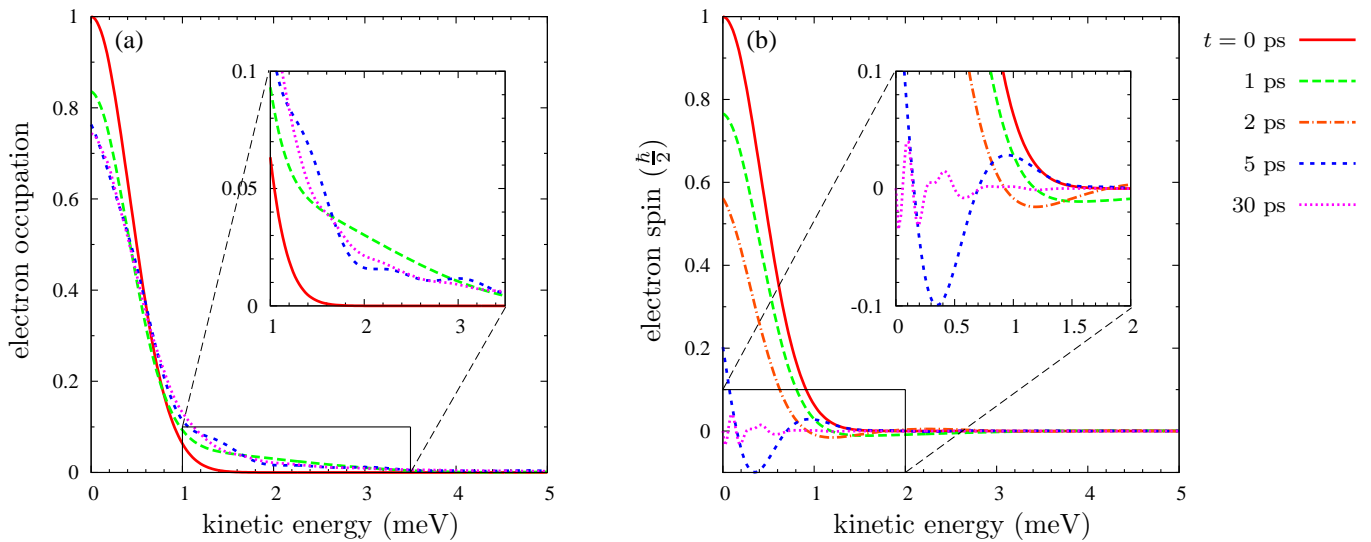}
\par\end{centering}
\caption{(color online) Distribution of (a) the electron occupation and (b) the electron spin as a function of the kinetic 
energy for different times $t$. The insets show magnified sections of the main figures.}
\label{fig:6} 
\end{figure*} 

In order to analyze these redistributions in more detail we have plotted in Fig.~\ref{fig:6}  (a) the electron and 
in Fig.~\ref{fig:6} (b) the spin distribution for different times $t$, i.e.
cross sections of Figs.~\ref{fig:4} and \ref{fig:5}, respectively. While the initial electrons are Gaussian distributed 
[cf. Eq.~\eqref{eq:8a}],
the final electron distribution is a Lorentzian curve in good approximation. The local minima and maxima in the electron 
occupation, which can be clearly identified in the inset of Fig.~\ref{fig:6} (a), correspond to the tail structure 
in Fig.~\ref{fig:4}.
Although the total electron spin is basically gone after 20 ps [cf. Fig.~\ref{fig:1} (b)], 
the spin distribution is not zero yet for all energies but rather quickly oscillates with respect to the energy
[cf. inset in Fig.~\ref{fig:6}(b)]. With increasing $t$,
these oscillations become fast and their amplitude slowly vanishes. Thus, 
eventually the electron spin distribution  approaches  zero which implies that the initial
electronic spin is completely transferred to the Mn subsystem. 

This behavior should be contrasted with spin assimilation processes described by rate equations
where the electron energy is conserved. In such a process the majority (minority) spin at a given energy 
decreases (increases) until any excess spin is gone. Thus, at each energy the total spin cannot switch its sign. 
In contrast, quantum kinetics allows that the majority spin  at a given energy is 
flipped and transferred to a different energy. 
Thus in certain regions of energy majority and minority spins switch roles repeatedly  and it depends on details
of the distribution whether or not the total electron spin switches sign. 
Indeed, the overshoot in quantum films and the oscillations in quantum wires demonstrate that such a reversal
of the sign of the total electron spin may actually take place [cf.~Fig.~\ref{fig:1}]. 
These observations suggest that the most severe approximation implied by replacing the 
quantum kinetic memory function by a Markovian rate is not the neglect of a finite memory time
but the suppression of the redistribution in energy space that in the quantum kinetic theory accompanies 
the electron-Mn scattering. 
In order to substantiate this assumption we have developed a level of description where the
memory depth is kept finite but the energy redistribution is artificially suppressed.
To this end  we make the ansatz
 \begin{align}
\label{eq:20}
\C{}{}{}(t) = \tilde C_\sigma^\sigma(t)f(\bk),
\end{align}
for the electron variables where we force the electron distribution $f(\bk)$ to be time independent
which implies that the energy distribution is kept fixed.
Without loss of generality  $f(\bk)$ can be assumed to be normalized according to
$$1 
%N%
= \int\limits_{BZ} f(\bk) \dkd.$$
Inserting this ansatz in Eqs.~\eqref{eq:11} 
and integrating Eq.~\eqref{eq:11b} over $\bk_{1}$
we obtain 
\begin{subequations}
\label{eq:21}
\begin{align}
\label{eq:21a}
&\dt \M{_1}{_1} 
=
\frac {2 J_{sd}^2L^{2D}} { \hbar^2 V^2(2\pi)^{2D}}  
\sum_{n\sigma\sigma'}  \big( \bS_{nn_1} \cdot  \bs_{\sigma\sigma'}  \big)^2    \nn
& \times\int\limits_{t_0}^t \bigg[
\Big( \tilde C_{\sigma}^{\sigma}(t')\M{}{}(t') - \tilde C_{\sigma'}^{\sigma'}(t')\M{_1}{_1}(t')
 \Big) G^D(t - t') dt' , \allowdisplaybreaks\\
\label{eq:21b}
&\dt  \tilde C_{\sigma_1}^{\sigma_1} 
= 
\frac {2J_{sd}^2  n_{Mn} L^D} { \hbar^2 V (2\pi)^D} \int\limits_{t_0}^t \sum_{nn'\sigma} 
 \big(\bS_{nn'} \cdot \bs_{\sigma\sigma_1} \big)^2 \nn
& \quad \times
 \Big(  \tilde C_{\sigma}^{\sigma} \M{}{} - \tilde C_{\sigma_1}^{\sigma_1} \M{'}{'}\Big) G^D(t-t')   dt'. % \frac 1 N
\end{align}
\end{subequations}
Here we have introduced the $\bk$-integrated memory function
\begin{align}
\label{eq:22}
G^D(\tau) 
%&= \iint\limits_{BZ} \cos\Big(\frac 1 \hbar\big(E_{\bk'}-E_{\bk}\big)\tau\Big) \dkd' \bigg]  f(\bk)\dkd \nn
&= \int\limits_{BZ}   G^D_\bk(\tau)   f(\bk)\dkd,
\end{align}
where $G^D_\bk(\tau)$ was defined in Eq.~\eqref{eq:12}.
Solving these equations for the example of a quantum well with the initial conditions  Eqs.~\eqref{eq:8}
[cf. pink line with triangles in Fig.~\ref{fig:1} (b)], 
essentially reproduces the results, we have previously obtained from the corresponding rate equations, 
even though we now account for the finite memory depth  and only neglect the redistribution of electron energies. 
In particular, the spin overshoot can only be described with the full quantum kinetic theory. 
It thus has to be concluded that these redistributions are indeed the actual source 
of the reversal of the sign of the electron spin observed in the quantum kinetic simulations
and that the neglect of the finite memory time is the less severe approximation when 
deriving the Markovian rate equations.

\begin{figure*}[tp]
\begin{centering}
\includegraphics[width=\textwidth]{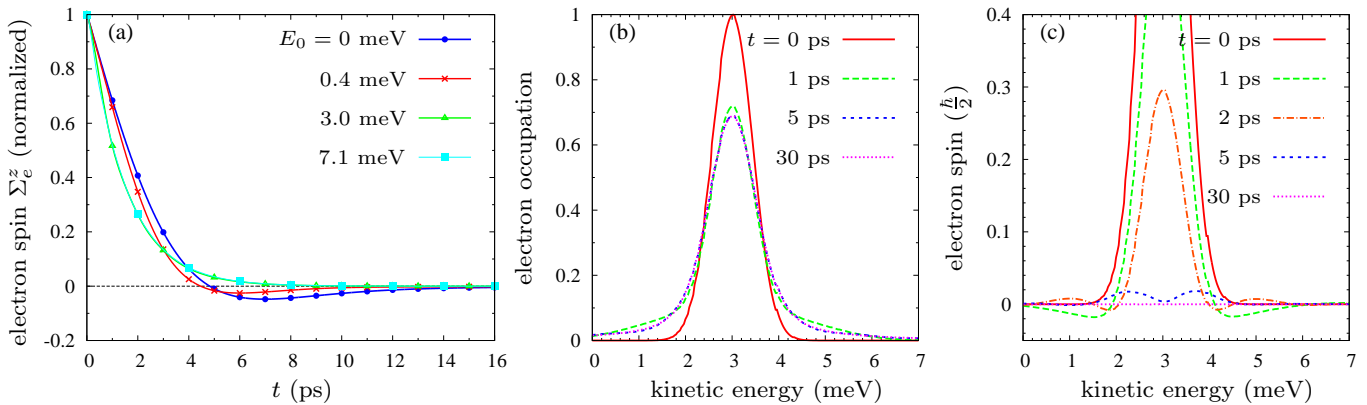}
\par\end{centering}
 \caption{(color online)
(a) Time evolution of the total electron spin $\Sigma_e^z$ for different central energies $E_0$
of the initial Gaussian distribution; 
(b) distribution of the electron occupation as a function of the kinetic 
energy for the central energy $E_0 = 3$ meV and different times $t$;
(c) the corresponding distribution of the electron spin
}
\label{fig:7} 
\end{figure*}

The same analysis for bulk and quantum wire systems reveals that energy redistribution of electrons 
and corresponding spin oscillations in the Brillouin zone are also present, 
but they are far less pronounced in the 3D case. 
Hence, the rate equations agree with the full quantum kinetic 
equations [cf. Fig.~\ref{fig:1} (a)] for bulk systems and predict a monotonic spin assimilation in this case. 
In quantum wire systems, on the other hand, these effects are 
much stronger, leading to an oscillating total electron spin [cf. Fig.~\ref{fig:1} (c)].
It is interesting to compare our present results also with the Rabi-type spin exchange observed in
single quantum dots doped with a single Mn atom.\cite{Reiter:09,Reiter:11,Reiter:12}
In a  Rabi-type spin transfer essentially two discrete states are involved that are coupled by the
exchange interaction. 
The dynamics of this two-level system
can be reduced to the dynamics of the
occupations by formally solving the equation of motion for  
the coherence between these states.
Inserting the result into the equations of motion for the occupations 
yields a memory function representing the coherence.
Noting, that in our case where the Mn concentration is much larger than the itinerant
electron concentration and therefore changes of $\M{}{}$ are negligible in 
determining the back action of the Mn atoms on the carriers,
we find according to Eq.~\eqref{eq:11b} 
that also in our case the change of the electronic occupations at time $t$ 
is determined essentially only by their values at earlier times.
However, in the two-level case there are only two occupations and these are not independent variables, 
because of the charge conservation. Thus, in this case a given occupation is coupled only to its own values at earlier times.
Therefore, the non-monotonic time evolution of the electronic spin as manifested in the corresponding Rabi-flops
can only take place when the memory provided by the coherence is long, as otherwise we would necessarily reach the
limit of Markovian rates. An energetic redistribution of carriers, as required for coherent spin transfer 
in quantum wells and wires, is not possible because there are no further final states
available in a two-level system. We therefore conclude that the mechanisms responsible 
for coherent spin exchange differ in extended semiconductors  qualitatively from those in quantum dots.

We note in passing that we have  performed a series of simulations for different shapes and widths of the initial distributions [not shown]
which lead to the conclusion that, as a general rule, the redistribution of electronic energies and the resulting effect on the
spin dynamics is more significant the sharper occupied and unoccupied regions are initially separated in the Brillouin zone. 
For example, relatively broad initial distributions which resemble a Lorentzian curve do not tend to show a spin overshoot, 
while the overshoot is especially pronounced for narrow box like 
distributions with steep edges.

Apart from the shape and width another important parameter 
of  the initial distribution is its energetic position determined 
by the central energy $E_{0}$ of the Gaussian electron distribution given in Eq.~\eqref{eq:8a}.
Fig.~\ref{fig:7} (a) shows the total electron spin as a function of time for different
values of $E_0$. The curve with $E_0 = 0$ meV has already been shown in Fig.~\ref{fig:1} (b) and is repeated here for better comparison.
For $E_0 = 0.4$ meV, which equals the standard deviation $\Delta$ of the Gaussian distribution, the spin overshoot
is less distinct. For $E_0 \gg \Delta$, the time evolution
of the total electron spin no longer depends on $E_0$ and the overshoot disappears. 
The electron and spin distribution for $E_{0}=3$ meV 
are depicted in Figs.~\ref{fig:7} (b) and (c), respectively.
It is seen that energetic redistributions take place also in this case and the spin distribution oscillates as a function
of time and energy. Compared with the corresponding distributions for $E_{0}=0$ meV [cf. Figs.~\ref{fig:6} (a) and (b)] 
these distributions have a higher symmetry and the excess spin initially prepared at the central energy $E_{0}$ decreases 
faster as now redistributions to lower and higher energies are possible. It turns out that these quantitative differences
altogether have the effect that here after summing over the individual spins the oscillations average out and the total
spin simply decays monotonically. 
A similar behavior is found in quantum well systems, where the oscillations of the total electron spin are
 most pronounced if the  initial electron distribution is in the vicinity of 
the band edge (not shown). This observation suggests that the different density of states at the band edge 
in bulk systems, quantum wells, and quantum wires is is an important factor for the appearance and strength of the coherent
phenomena in the spin transfer.

\section{Conclusion}

We have analyzed the spin transfer from an initially prepared electronic excess spin towards the spin of Mn atoms in
a diluted Mn doped magnetic semiconductor within a quantum kinetic theory.
We have demonstrated that these spin transfer processes are dominated by only a few relevant
correlations which form a small subset of the full set of terms that contribute to the quantum
kinetic equations on this level of the correlation expansion. By concentrating  on these terms only,
it is possible to eliminate the pertinent correlations in favor of a memory function. Although
the typical memory times are orders of magnitude shorter than commonly encountered 
spin transfer times, noticeable deviations from  Markovian dynamics
may occur for quantum well and wire systems while for bulk systems we find a good agreement between
quantum kinetic calculations and the Markovian limit.
For wells and wires  the spin transfer involves one or more changes of the sign of the total electron spin
which indicates a regime of coherent exchange of spin between the electronic and the Mn subsystems.
In contrast to Rabi-type spin-exchange in Mn doped quantum dots, these coherent dynamical features are here not
related to a long memory time. Instead, it is a redistribution of electronic energies due to the
energy-time uncertainty which enables a spin-exchange where the total electron spin exhibits a non-monotonic
time evolution. Markovian rate equations fail to describe this type of dynamics because
the energy-time uncertainty is suppressed in this limit and thus the necessary energetic
redistributions do not take place.

\begin{appendix}
 \section{Source terms for correlations} 
In this appendix we give the explicit form of the source term $Q_{\bar K}$ on the right hand side of 
the equations of motion \eqref{eq:5d} for the correlations $\bar K$. It reads:
\begin{widetext}
\label{eq:6}
\begin{align}
\label{eq:6a}
\!\!\!&Q_{\K{_1}{_2}{_1}{_2}{_1}{_2}} =   
\!J_{sd} \Bigg\{ 
 \int\limits_{BZ} \!\!\bigg[ {\bS}_{n_2n_1}\!\! \cdot  {\bs}_{ \sigma_2 \sigma_{1}}\! \Big(\!    
\Ck{_2}{_2}{}{_2} \M{_2}{_2} \!-\! \Ck{_1}{_1}{_1}{} \M{_1}{_1}\!\Big)
\! + \!  \sum_{n \sigma}  \! \Big(
  {\bS}_{nn_1} \!\!\cdot  {\bs}_{ \sigma \sigma_{1}} 
\K{}{_2}{}{_2}{}{_2}
\!  - {\bS}_{n_2n}\! \cdot  {\bs}_{ \sigma_2 \sigma} 
 \K{_1}{}{_1}{}{_1}{}
\!\Big)\!\bigg] 
\frac {L^D\dkd } {V(2\pi)^D}  
%%%%%%%%%%%%%% 4
\nn
&
+\!
 \sum_{n} \!
\Bigg[ \M{}{}  {\bS}_{nn}\! \cdot\! \bigg[
 n_{\text{Mn}}   \K{_1}{_2}{_1}{_2}{_1}{_2}  \Big(  {\bs}_{ \sigma_1 \sigma_1}\!\! - {\bs}_{ \sigma_2 \sigma_2} \! \Big) 
\!-  \delta_{n_1n_2}\delta_{\sigma_1\sigma_2} \M{_1}{_1} {\bs}_{ \sigma_1 \sigma_1}\! \Big(   \C{_1}{_1}{_2}\!
 -  \C{_1}{_1}{_1} \!
 +  \!  \int\limits_{BZ} \!\!
  \big(   \Ck{_1}{_1}{}{_2} 
-    \Ck{_1}{_1}{_1}{} \big)  \frac {L^D\dkd}{V(2\pi)^D}   \Big) \!\bigg]
\nn &
- \delta_{n_1n_2} \sum_{n' \sigma} \M{_1}{_1} \!  \int\limits_{BZ}
 {\bS}_{nn'} \!\cdot\!  \Big(  {\bs}_{ \sigma \sigma_1} \K{}{_2}{}{'}{}{_2} 
-   {\bs}_{ \sigma_2 \sigma} \K{_1}{}{}{'}{_1}{} \Big) \frac {L^D\dkd}{V(2\pi)^D} 
\Bigg]
\! -\!     
 \int\limits_{BZ} \sum_{\sigma } \bigg[   
 {\bs}_{ \sigma \sigma} \cdot  \Big({\bS}_{n_2n_2} 
 \!-  {\bS}_{n_1n_1}     \Big) \C{}{}{} \K{_1}{_2}{_1}{_2}{_1}{_2} 
\nn
& -
\sum_n {\bs}_{ \sigma \sigma_1} \cdot 
   \Big({\bS}_{nn_1}  \K{}{_2}{}{_2}{}{_2}  - {\bS}_{n_2n}  \K{}{_2}{_1}{}{}{_2} \Big)  \C{_1}{_1}{_1}
  -   \sum_{n} {\bs}_{ \sigma_2 \sigma} \cdot  
 \Big({\bS}_{nn_1} 
\K{_1}{}{}{_2}{_1}{}  - 
{\bS}_{n_2n}  \K{_1}{}{_1}{}{_1}{} \Big) \C{_2}{_2}{_2} 
\bigg]  \frac {L^D\dkd}{V(2\pi)^D} \nn
& -  {\bs}_{ \sigma_2 \sigma_1} \cdot {\bS}_{n_2n_1}   \Big(
 \M{_2}{_2}  -  \M{_1}{_1} \Big) \bigg[ \int\limits_{BZ}   \Big( \C{_1}{_1}{_1} \Ck{_2}{_2}{}{_2} +  
\C{_2}{_2}{_2} \Ck{_1}{_1}{_1}{} \Big) \frac {L^D\dkd}{V(2\pi)^D}   
+
 \C{_2}{_2}{_2} \C{_1}{_1}{_1} \bigg]
 \nn
& - \delta_{\sigma_1\sigma_2}\Ck{_1}{_1}{_1}{_2}\iint\limits_{BZ}  \sum_{n \sigma \sigma'} {\bs}_{ \sigma \sigma'} 
\cdot \Big( {\bS}_{nn_1}  
 \K{}{'}{}{_2}{}{'}  - {\bS}_{n_2n}  \K{}{'}{_1}{}{}{'} \Big)  \frac {L^{2D} \dkd' \dkd}{V^2(2\pi)^{2D}}  \Bigg\} .
\end{align}
\end{widetext}
\end{appendix}

\end{document}